\documentclass[11pt]{article}
\usepackage{mathrsfs}
\usepackage{euscript}

\def \beq{\begin{equation}}
\def \eeq{\end{equation}}
\def \beqa{\begin{eqnarray}}
\def \eeqa{\end{eqnarray}}

\input{tcilatex}

\begin{document}

\author{ \textsc{Adam Bzdak} 
\\
Institute of Physics, Jagiellonian University,
\\ 
Reymonta 4, 30-059 Krak\'ow, Poland \\
E-mail: \texttt{bzdak@th.if.uj.edu.pl} \\
\\
\textsc{Jerzy Szwed} \footnote {Work supported by
the European Commission contract ICA1-CT-2002-70013/INCO Strategic action on training and excellence, 5FP}
\\
Institute of Physics, Jagiellonian University,
\\ 
Reymonta 4, 30-059 Krak\'ow, Poland 
\\and
\\
Centre de Physique Th\'eorique, CNRS Luminy Case 907, \\
13288 Marseille Cedex 09, France \\
E-mail: \texttt{szwed@cpt.univ-mrs.fr} }
\title{\small CENTRE DE PHYSIQUE TH\'EORIQUE \footnote{Unit\'e Mixte de Recherche du CNRS et des Universit\'es 
de Provence, de la M\'editerran\'ee et du Sud Toulon-Var - Laboratoire affili\'e a la FRUNAM - FR 2291}
\\
CNRS-Luminy, Case 907
\\
13288 Marseille Cedex 9
\\
FRANCE
\vskip0.5cm
\large \bf THE 'SQUARE ROOT' OF THE INTERACTING DIRAC EQUATION}
\maketitle
\vskip 0mm

\noindent CPT-2004/P....
\\
\noindent TPJU-11/2004

\begin{abstract}
The 'square root' of the interacting Dirac equation is constructed. The obtained equations lead 
to the Yang-Mills superfield with the appropriate equations of motion for the component fields.
\end{abstract}

\newpage

\textbf{1. Introduction.}

The idea of taking the square root of the Dirac equation was originally
proposed in Ref. \cite{Szwed}. It was shown that natural scene for such
construction is superspace, and the new field equations were proposed. Such
idea follows directly from the analogous procedure performed by Dirac on the
Klein-Gordon equation \cite{Dirac}.

The construction was re-considered in Ref. \cite{BzdakHadasz} and a set of
new free field equations was solved. The solution turned out to be the
Maxwell superfield with appropriate equations of motion for the component
fields $i.e.$ the Maxwell equations (the ''photon'') and the massless Dirac
equation (the ''photino'').

In the present paper we demonstrate how to construct the square root of the
interacting (with non-abelian vector gauge field) Dirac equation. We also
construct the corresponding field equations and study their solutions.

\textbf{2. Solution.}

Let us define the covariant derivative:%
\begin{equation}
{\mathscr D}_{\mu }=\partial _{\mu }+ig\hat{A}_{\mu },
\end{equation}%
and write the Dirac equation coupled to the vector gauge field $\hat{A}_{\mu
}\equiv A_{\mu }^{a}(x)T^{a}$ (with the Hermitian matrices $T^{a}$ being the
gauge group generators) as:%
\begin{equation}
\left( i\gamma ^{\mu }{\mathscr D}_{\mu }-m\right) \Psi =0.
\end{equation}%
Using the two-component notation in the chiral representation one can
rewrite this equation as:%
\begin{equation}
-\left( 
\begin{array}{cc}
i\bar{\sigma}^{\mu \,\dot{\alpha}\alpha }{\mathscr D}_{\mu } & m \\ 
m & i\sigma ^{\mu }{}_{\alpha \dot{\alpha}}{\mathscr D}_{\mu }%
\end{array}%
\right) \left( 
\begin{array}{c}
\varphi _{\alpha } \\ 
\bar{\chi}^{\dot{\alpha}}%
\end{array}%
\right) \equiv \mathcal{D}_{int}\left( 
\begin{array}{c}
\varphi _{\alpha } \\ 
\bar{\chi}^{\dot{\alpha}}%
\end{array}%
\right) \;=\;0.
\end{equation}%
The Lorenz indices are denoted here by $\mu $, $\nu $, $\lambda $ and $\rho
, $ the spinor indices by $\alpha $ and $\beta $.

We are now looking for the operator $\mathcal{S}_{int}$ satisfying:%
\begin{equation}
\mathcal{S}_{int}^{\dagger }\mathcal{S}_{int}=\mathcal{D}_{int}.
\label{S+S=D}
\end{equation}%
The solution is analogical to the non-interacting case when expressed in
terms of the covariant spinorial derivatives:%
\begin{equation}
{\mathscr D}_{\alpha }=\left( \partial /\partial \theta ^{\alpha }+i\sigma
^{\mu }{}_{\alpha \dot{\alpha}}\bar{\theta}^{\dot{\alpha}}\partial _{\mu
}\right) +ig\hat{A}_{\alpha },\quad \hat{A}_{\alpha }\equiv A_{\alpha
}^{a}T^{a},\quad \bar{{\mathscr D}}_{\dot{\alpha}}=\left( {\mathscr D}%
_{\alpha }\right) ^{\star },
\end{equation}%
and reads:%
\begin{equation}
\mathcal{S}_{int}={\frac{1}{\sqrt{2}}}\left( 
\begin{array}{rr}
{\mathscr D}^{\alpha } & -\bar{{\mathscr D}}_{\dot{\alpha}} \\ 
\bar{{\mathscr D}}^{\dot{\alpha}} & {\mathscr D}_{\alpha }%
\end{array}%
\right) .  \label{Sint_def}
\end{equation}%
Indeed, using the anticomutation relation:%
\begin{equation}
\left\{ {\mathscr D}_{\alpha },\bar{{\mathscr D}}_{\dot{\alpha}}\right\}
=-2i\sigma ^{\mu }{}_{\alpha \dot{\alpha}}{\mathscr D}_{\mu }
\end{equation}%
we reproduce the Dirac operator $\mathcal{D}_{int}$ with, in analogy to the
free field case, an operator $M_{int}$ appearing in place of the mass $m$.
In the present case it is defined as:%
\begin{equation}
M_{int}=-\frac{1}{4}\left( {\mathscr D\mathscr D}+\bar{{\mathscr D}}\bar{{%
\mathscr D}}\right) .
\end{equation}%
We conclude that the operator $\mathcal{S}_{int}$ is the square root of the
interacting Dirac operator $\mathcal{D}_{int}$ when acting within the space
of fields $\Lambda $ fulfilling the 'mass condition' \cite{BzdakHadasz}:%
\begin{equation}
M_{int}\Lambda =m\Lambda .
\end{equation}%
On this space we can therefore consider the equation:%
\begin{equation}
\mathcal{S}_{int}\Lambda =0  \label{Sint_L=0}
\end{equation}%
as a square root of the interacting Dirac equation.

Following \cite{Szwed} and \cite{BzdakHadasz} we are interested in the
solutions of Eq. (\ref{Sint_L=0}) having the form:%
\begin{equation}
\Lambda =\left( 
\begin{array}{r}
W_{\alpha } \\ 
\bar{\mathcal{H}}^{\dot{\alpha}}%
\end{array}%
\right) .
\end{equation}%
Using Eq. (\ref{Sint_def}) we obtain the following set of equations:%
\begin{equation}
\mathcal{S}_{int}\left( 
\begin{array}{r}
W_{\alpha } \\ 
\bar{\mathcal{H}}^{\dot{\alpha}}%
\end{array}%
\right) ={\frac{1}{\sqrt{2}}}\left( 
\begin{array}{rr}
{\mathscr D}^{\alpha } & -\bar{{\mathscr D}}_{\dot{\alpha}} \\ 
\bar{{\mathscr D}}^{\dot{\alpha}} & {\mathscr D}_{\alpha }%
\end{array}%
\right) \left( 
\begin{array}{r}
W_{\alpha } \\ 
\bar{\mathcal{H}}^{\dot{\alpha}}%
\end{array}%
\right) =0
\end{equation}%
together with:%
\begin{equation}
M_{int}\left( 
\begin{array}{r}
W_{\alpha } \\ 
\bar{\mathcal{H}}^{\dot{\alpha}}%
\end{array}%
\right) =m\left( 
\begin{array}{r}
W_{\alpha } \\ 
\bar{\mathcal{H}}^{\dot{\alpha}}%
\end{array}%
\right) .  \label{rowMexp}
\end{equation}%
Note that due to Eq. (\ref{S+S=D}) the above superfields obey also the Dirac
equation. Using the techniques analogous to those of Ref. \cite{BzdakHadasz}
we obtain:%
\begin{equation}
{\mathscr D}^{\beta }\left( \bar{{\mathscr D}}_{\dot{\alpha}}W_{\alpha
}\right) =\bar{{\mathscr D}}^{\dot{\beta}}\left( \bar{{\mathscr D}}_{\dot{%
\alpha}}W_{\alpha }\right) ={\mathscr D}_{\mu }\left( \bar{{\mathscr D}}_{%
\dot{\alpha}}W_{\alpha }\right) =0,  \label{3row1}
\end{equation}%
\begin{equation}
{\mathscr D}^{\beta }\left( {\mathscr D}_{\alpha }\bar{\mathcal{H}}^{\dot{%
\alpha}}\right) =\bar{{\mathscr D}}^{\dot{\beta}}\left( {\mathscr D}_{\alpha
}\bar{\mathcal{H}}^{\dot{\alpha}}\right) ={\mathscr D}_{\mu }\left( {%
\mathscr D}_{\alpha }\bar{\mathcal{H}}^{\dot{\alpha}}\right) =0.
\label{3row2}
\end{equation}%
Let us look closer at the last equality in Eq. (\ref{3row2}). Multiplying it
by $\partial _{\nu }$ and subtracting from $\partial _{\mu }{\mathscr D}%
_{\nu }\left( \bar{{\mathscr D}}_{\dot{\alpha}}W_{\alpha }\right) $ we
obtain:%
\begin{equation}
\hat{F}_{\mu \nu }\left( \bar{{\mathscr D}}_{\dot{\alpha}}W_{\alpha }\right)
=0,
\end{equation}%
where $\hat{F}_{\mu \nu }$ is the gauge field tensor:%
\begin{equation}
\hat{F}_{\mu \nu }=\partial _{\mu }\hat{A}_{\nu }-\partial _{\nu }\hat{A}%
_{\mu }+ig\left[ \hat{A}_{\mu },\hat{A}_{\nu }\right] .  \label{defF}
\end{equation}%
The case $\hat{F}_{\mu \nu }=0$ corresponds to the free field system and was
solved in Ref. \cite{BzdakHadasz}. With the gauge interactions switched on
we have thus:%
\begin{equation}
\bar{{\mathscr D}}_{\dot{\alpha}}W_{\alpha }=0.  \label{eq1}
\end{equation}%
In analogous way we obtain:%
\begin{equation}
{\mathscr D}_{\alpha }\bar{\mathcal{H}}^{\dot{\alpha}}=0.  \label{eq2}
\end{equation}%
The mass condition can be now simplified. Multiplying the upper Eq. (\ref%
{rowMexp}) by ${\mathscr D}^{\alpha }$ we obtain:%
\begin{equation}
m{\mathscr D}^{\alpha }W_{\alpha }=0.
\end{equation}%
Out of two possible choices let us first assume $m\neq 0,$ ${\mathscr D}%
^{\alpha }W_{\alpha }=0.$ We use the Dirac equation to prove that:%
\begin{equation}
-2i\bar{\sigma}^{\mu \,\dot{\alpha}\alpha }{\mathscr D}_{\mu }W_{\alpha
}=\left\{ {\mathscr D}^{\alpha },\bar{{\mathscr D}}^{\dot{\alpha}}\right\}
W_{\alpha }=2m\bar{\mathcal{H}}^{\dot{\alpha}}=0,
\end{equation}%
and thus $\bar{\mathcal{H}}^{\dot{\alpha}}=0.$ Using the lower Eq. (\ref%
{rowMexp}) and the second component of the Dirac equation we arrive
analogically at $W_{\alpha }=0.$ Therefore the nontrivial solutions to our
system require the mass to vanish:%
\begin{equation}
m=0.
\end{equation}%
As an immediate consequence Eqs. (\ref{rowMexp}) reduce to:%
\begin{equation}
{\mathscr D}^{2}W_{\alpha }=\bar{{\mathscr D}}^{2}\bar{\mathcal{H}}^{\dot{%
\alpha}}=0.
\end{equation}%
Repeating analogous considerations which have led to Eqs. (\ref{eq1}, \ref%
{eq2}) we obtain:%
\[
{\mathscr D}^{\alpha }W_{\alpha }=\bar{{\mathscr D}}_{\dot{\alpha}}\bar{W}^{%
\dot{\alpha}}=0, 
\]%
\begin{equation}
\bar{{\mathscr D}}_{\dot{\alpha}}\bar{\mathcal{H}}^{\dot{\alpha}}={\mathscr D%
}^{\alpha }\mathcal{H}_{\alpha }=0.
\end{equation}%
The above equations together with the equations (\ref{eq1}), (\ref{eq2})
define the supermultiplets $W_{\alpha }$ and $\bar{\mathcal{H}}^{\dot{\alpha}%
}$ and their dynamic equations. The solution has the form of the Yang-Mills
superfield \cite{WessBagger}\cite{Weinberg}:%
\begin{eqnarray}
W_{\alpha } &=&-i\hat{\lambda}_{\alpha }\left( y\right) +\left[ \delta
_{\alpha }{}^{\beta }\hat{d}\left( y\right) -\frac{i}{2}\left( \sigma ^{\mu }%
\bar{\sigma}^{\nu \,}\right) _{\alpha }{}^{\beta }\hat{F}_{\mu \nu }\left(
y\right) \right] \theta _{\beta }  \nonumber \\
&&+\theta \theta \sigma ^{\mu }{}_{\alpha \dot{\alpha}}{\mathscr D}_{\mu }%
\hat{\bar{\lambda}}^{\dot{\alpha}}(y),
\end{eqnarray}%
with:%
\begin{eqnarray}
y^{\mu } &=&x^{\mu }+i\theta \sigma ^{\mu }\bar{\theta}, \\
\hat{\lambda}_{\alpha }\, &=&\,\lambda _{\alpha }^{a}\,T^{a}\,,\quad \hat{d}%
\,=\,d^{a}\,T^{a},  \nonumber \\
{\mathscr D}_{\mu }\hat{\bar{\lambda}}^{\dot{\alpha}}\, &=&\,\partial _{\mu }%
\hat{\bar{\lambda}}^{\dot{\alpha}}-ig\,[\hat{A}_{\mu },\hat{\bar{\lambda}}^{%
\dot{\alpha}}],  \nonumber
\end{eqnarray}%
and analogous expression for the field $\bar{\mathcal{H}}^{\dot{\alpha}}.$
The fields $\hat{F}_{\mu \nu }$, $\hat{\lambda}_{\alpha }$ and $\hat{d}$
satisfy corresponding dynamic equations:%
\begin{eqnarray*}
{\mathscr D}_{\mu }\hat{F}^{\mu \nu } &=&0,\qquad \epsilon ^{\mu \nu \lambda
\rho }{\mathscr D}_{\nu }\hat{F}_{\lambda \rho }=0, \\
\sigma ^{\mu }{}_{\alpha \dot{\alpha}}{\mathscr D}_{\mu }\hat{\bar{\lambda}}%
^{\dot{\alpha}} &=&0,\qquad \hat{d}=0.
\end{eqnarray*}

\textbf{3. Conclusions.}

In this paper we show how to define the square root of the Dirac equation
coupled to the non-abelian gauge field. The obtained equations lead to the
Yangs-Mills superfield with appropriate equations of motion for the
component fields.

It may be interesting to extend the results of this paper in several
directions.

It is worth studying others form of solutions, proposed in Ref. \cite{Szwed}%
. This problem is currently under investigation.

Starting from the Dirac equation on the curved space-time, one may hope to
obtain some form of supergravity.

It seems also interesting to extend our consideration to higher space-time
dimensions.


\bigskip

\begin{thebibliography}{9}
\bibitem{Szwed} J. Szwed, Phys. Lett. \textbf{B 181}, 305, (1986).

\bibitem{Dirac} P. A.~M.~Dirac, Proc.\ R.\ Soc.\ (London) \textbf{A 117},
610, (1928); \textbf{A 118}, 351, (1928).

\bibitem{BzdakHadasz} A. Bzdak, L. Hadasz, Phys. Lett. \textbf{B 582}, 113,
(2004).

\bibitem{WessBagger} J.~Wess, J.~Bagger, \textsl{Supersymmetry and
Supergravity,} Princeton University Press, Princeton, 1983.

\bibitem{Weinberg} S.~Weinberg, \textsl{The Quantum Theory of Fields. Vol.
III Supersymmetry,} Cambridge University Press, Cambridge, 2000.
\end{thebibliography}
\end{document}